\documentclass[12pt]{article}
\usepackage{amssymb,amsmath,amsthm,amscd,latexsym}
\usepackage{mathrsfs}
\usepackage{mathrsfs}
\usepackage{amsfonts}
\usepackage{amsmath}
\usepackage{amssymb}
\usepackage{amscd}
\usepackage{mathrsfs}
\usepackage{amssymb}
\usepackage{amsmath}
\usepackage{amsthm}
\usepackage{latexsym}
\usepackage{indentfirst}
\usepackage{enumitem}
\usepackage{anysize}
\usepackage{bbm}

\renewcommand{\paragraph}{\roman{paragraph}}
\input cyracc.def

\newcommand{\F}{\mathbb{F}}

\newtheorem{theorem}{Theorem}

\theoremstyle{definition}

\begin{document}
\title{\bf Two-weight codes and second order recurrences
\thanks{This research is supported by National Natural Science Foundation of China (61672036),
Technology Foundation for Selected Overseas Chinese Scholar, Ministry of Personnel of China (05015133) and
the Open Research Fund of National Mobile Communications Research Laboratory, Southeast University (2015D11) and
Key projects of support program for outstanding young talents in Colleges and Universities (gxyqZD2016008).}}

\author{
\small{Minjia Shi$^{1,2}$, Zhongyi Zhang$^4$, and Patrick Sol\'e$^5$}\\ 
\and \small{${}^1$Key Laboratory of Intelligent Computing \& Signal Processing,}\\
 \small{Ministry of Education, Anhui University No. 3 Feixi Road,}\\
  \small{Hefei Anhui Province 230039, P. R. China;}\\
\small{${}^2$School of Mathematical Sciences, Anhui University, Hefei, 230601, China}\\
\small {${}^3$National Mobile Communications Research Laboratory}\\
\small {Southeast University, 210096, Nanjing,  P. R. China;}\\
[-0.8ex]
\small{${}^4$School of Wendian, Anhui University, Hefei, 230601, P. R. China}\\
\small{${}^5$CNRS/LAGA, University of Paris 8, 2 rue de la Libert\'e, 93 526 Saint-Denis, France}}
\date{}
\maketitle
\begin{abstract} Cyclic codes of dimension $2$ over a finite field are shown to have at most two nonzero weights. This extends a construction of Rao et al (2010).
We compute their weight distribution, and give a condition on the roots of their check polynomials for them to be MDS.
\end{abstract}


\section{Introduction}
The topic of two-weight codes and their many connections with strongly regular graphs, association schemes \cite{BH}, and finite geometries \cite{Cald}, has been explored since the seminal work of Delsarte \cite{D}. To construct two-weight codes, an old technique is to use irreducible cyclic codes, which, in turn can be studied by Gauss sums, Fourier transform and L-series \cite{M}.  A family of two weight cyclic codes consisting of codes of dimension $2$ was found in \cite{RP}, and more recently in \cite{SZS}.

In the present paper, we generalize the results of \cite{RP,SZS} to show that all cyclic codes of dimension $2$ have at most two weights. Our techniques are very old in Number Theory \cite{L}, and very new in Coding Theory.
We extend the setting of \cite{R2} from prime fields to arbitrary finite fields, and of \cite{SZS} from the Fibonacci recurrence to an arbitrary second order recurrence. We consider the periodicity properties of zeroes in second order recurrence. There are three cases to consider, according to the factorization of
the check polynomial. In particular, the case when this polynomial is a square was not treated in \cite{SZS}. Our methods are elementary, and use little more than the form of the solution of a linear recurrence, depending on the factorization of its characteristic polynomial, as can be learned from, for instance, \cite{GKP}.

The material is organized as follows. The next section collects some technical prerequisites necessary to the following sections. Section 3 is central and derives the main results. Section 4 computes the weight distribution
of the two-weight codes constructed in the previous section. Section 5 is dedicated to numerical examples. Section 6 concludes the paper.
\section{Background material}
\subsection{Linear codes}
A {\bf (linear) code} $C$ of length $N$  over a finite field $\mathbb{F}_{q}$ is a $\mathbb{F}_{q}$ vector subspace of  $\mathbb{F}_{q}^n.$ The dimension of the code, is its dimension as a $\mathbb{F}_{q}$ vectorspace, and is denoted by $k.$ The elements of $C$ are called {\bf codewords}.

The {\bf dual} $C^ \bot$ of a code $C$ is understood w.r.t. the standard inner product.

 The {\bf (Hamming) weight} of $x \in \F_q^N$ is the number of indices $i$ where $x_i\neq 0.$ A code is {\bf two-weight} if it has only two non-zero weights amongst the weights of its codewords. The minimum nonzero weight $d$ of a linear code is called the {\bf minimum distance}. The {\bf dual distance} of a code is the minimum distance of its dual. A code is {\bf projective} if its dual distance is $\ge 3.$

 Every linear code satisfies the {\bf Singleton bound} \cite[Th 11, Chap. 1]{MS} on its parameters
 $$d\le n-k+1.$$
 A code meeting that bound is called MDS. See \cite[ Chap. 11]{MS} for general knowledge on this family of codes.
\subsection{Cyclic codes}
A {\bf cyclic} code of length $N$ over a finite field $\mathbb{F}_{q}$ is a $\mathbb{F}_{q}$ linear code of length $N$ invariant under the coordinate shift. Under the polynomial correspondence such a code can be regarded as an ideal in the ring $\mathbb{F}_{q}[x]/(x^N-1).$ It can be shown that this ideal is principal, with a unique monic generator $g(x),$ called {\bf the generator polynomial} of the code. The {\bf check polynomial} $h(x)$ is then defined as the
 quotient $(X^N-1)/g(x).$ A well-known fact is that the codewords are the periods of the linear recurrence of characteristic polynomial the reciprocal polynomial of $h(x)$ \cite[p. 195]{MS}.
 A cyclic code is {\bf irreducible} if its check polynomial $h(x)$ is. A code is two-weight if it has only two non-zero weights. In \cite{SW} a conjectural classification scheme of irreducible cyclic two-weight codes is given as
 \begin{enumerate}
 \item a list of eleven exceptional codes
 \item subfield codes
 \item semiprimitive irreducible codes
 \end{enumerate}
 {\bf Subfield codes} is the case when the root of $h(x)$ is a primitive root of a subfield of the quotient field $\mathbb{F}_{q}[x]/(h(x)).$
To define semiprimitive codes write $q^m-1=N u,$ where $m$ is the dimension of the code. If $-1$ is a power of $q$ modulo $u,$ then the irreducible cyclic code of parameters $[N,m]$ is said to be {\bf semiprimitive}. In this paper, we will exhibit several codes that do not fit this classification.
\section{Main results}
Assume a linear recurrence with characteristic polynomial $P(a,b;x)=x^2-ax-b$, with $a,b\in \F_q,\, b\neq 0,$ and period $N.$ Denote by $C(a,b,q)$ the cyclic code of length $N$  over $\F_q$ with check polymomial the reciprocal
of $x^2-ax-b$ that is $x^2+\frac{a}{b}x+\frac{1}{b}.$ 
In the special case of $q=p,$ and $p$ an odd prime, it is known in the literature of second order recurrences \cite{L,R2} that the zeros of a sequence defined by such a recurrence appear with periodicity $e$, say, and that $e$ divides $N$. The parameter
$e$ is called the {\bf rank} of the recurrence in \cite{R,R2}. We rederive the existence of such an $e$ in the next three subsections, each time with an algebraic characterization.

We call $g_n$ the generic solution of such a recurrence, with attached codeword $z=(g_0,\cdots,g_{N-1}),$ and we think of the weight of $z$ as $N-\vert \{n=0,1,\cdots,N-1 \mid g_n=0\}\vert.$
Thus, this weight is either $N$ or $N-\frac{N}{e}.$
\subsection{Irreducible cyclic code}
Assume a characteristic polynomial $P(a,b;x)=x^2-ax-b$ that is irreducible over $\F_q.$ Write $x^2-ax-b=(x-\alpha)(x-\beta),$ with $\alpha,\,\beta \in \F_{q^2}\setminus \F_q.$
{\theorem Let $P(a,b;x)$ be irreducible over $\F_q,$ with roots $\alpha$ and $\beta=\alpha^q.$ Let $K=\frac{N}{e},$ where $e$ is the order of $\frac{\beta}{\alpha}$ in $\F_{q^2}^*.$ \\
If $e<q+1,$ then the code $C(a,b,q)$ is a two-weight code with the two nonzero weights $\{N-K,\, N\}.$\\
If $e=q+1,$ then the code $C(a,b,q)$ is a one-weight code with the nonzero weight $N-K.$

   }
\begin{proof} Write
$$g_n=\lambda \alpha^n+\mu \beta^n=\alpha^n(\lambda+\mu (\frac{\beta}{\alpha})^n),$$ for some $\lambda, \, \mu \in \F_{q^2}.$
Since $g_n\in \F_q,$ and $\beta=\alpha^q,$ we have $\mu=\lambda^q.$ To avoid the trivial codeword assume $\lambda\neq 0.$
So $g_n=0$ is an equation in $n$ with either no solutions (if $-\frac{\lambda}{\mu}\notin \langle \frac{\beta}{\alpha} \rangle$) or $K$ solutions (if $-\frac{\lambda}{\mu}\in \langle \frac{\beta}{\alpha} \rangle$).
Since the value set of $-\frac{\lambda}{\mu}=-\frac{\lambda}{\lambda^q}$ has size $q+1$ the first case can only occur if $e<q+1.$
Note that, by definition, $e$ divides $N=ord(\alpha)=ord(\beta).$
\end{proof}
\subsection{Reducible cyclic code}
\subsubsection{Two roots}
Assume a characteristic polynomial $P(a,b;x)=(x-\alpha)(x-\beta),$ with $\alpha,\,\beta \in \F_q,$ and $\alpha \neq \beta.$ Write
$$g_n=\lambda \alpha^n+\mu \beta^n=\alpha^n(\lambda+\mu (\frac{\beta}{\alpha})^n),$$ for some $\lambda, \, \mu \in \F_q.$

{\theorem If $P(a,b;x)$ is reducible with two distinct roots over $\F_q,$ then the code $C(a,b,q)$ is a two-weight code with the two nonzero weights $\{N-K,\, N\}.$ Here $K=N/e$ where $e$ is the order of $\frac{\beta}{\alpha}$ in $\F_{q}^*.$ }
\begin{proof}
We discuss on the values of $(\lambda,\mu)\neq (0,0).$
\begin{itemize}
\item If  $ \mu$ is zero and $\lambda \neq 0,$ or $ \mu \neq 0 $ and $\lambda = 0,$ then $g_n \neq 0$ for all $0\le n\le N-1$.
\item If $\lambda \mu \neq 0,$ then
$$g_n=\mu\alpha^n\big(\frac{\lambda}{\mu}+ (\frac{\beta}{\alpha})^n\big),$$
an equation in $n$ with either no solutions (if $-\frac{\lambda}{\mu}\notin \langle \frac{\beta}{\alpha} \rangle$) or $K$ solutions (if $-\frac{\lambda}{\mu}\in \langle \frac{\beta}{\alpha} \rangle$).
\end{itemize}
Note that, by definition, $e$ divides $N=LCM(ord(\alpha),ord(\beta)).$
\end{proof}
\subsubsection{Single root}
Assume a characteristic polynomial $P(a,b;x)=(x-\alpha)^2,$ with $\alpha \in \F_q.$ The recurrence becomes $$g_{n+2}=2\alpha g_{n+1}-\alpha^2 g_n.$$ It is easy to check that the general solution is
$$g_n=\lambda \alpha^n+\mu n \alpha^n=\alpha^n(\lambda+\mu n),$$ for some $\lambda, \, \mu \in \F_q$ (See \cite{GKP} p. 341, with $\ell=1, \, d_1=2$).

{\theorem \label{square} If $P(a,b;x)$ is a square then the code $C(a,b,q)$ is a two-weight code with the two nonzero weights $N-\frac{N}{p},\, N.$ Here $p$ denotes the characteristic of $\F_q.$ }
\begin{proof}
We discuss on the values of $(\lambda,\mu)\neq (0,0).$
\begin{itemize}
\item If  $ \mu$ is zero and $\lambda \neq 0,$ then $g_n \neq 0.$
\item If  $ \lambda$ is zero and $\mu \neq 0,$ then $g_n =0$ whenever $p \vert n.$
 \item If $\lambda \mu \neq 0,$ then
$$g_n=\mu\alpha^n(\frac{\lambda}{\mu}+ n),$$
an equation in $n$ with $N/p$ solutions the $n$'s in the range $0\le n\le N-1$ such that $n\equiv -\frac{\lambda}{\mu}\pmod{p}.$
Note that, by Theorem \ref{fonda}, $p$ divides $N.$
\end{itemize}
Thus we prove the results.
\end{proof}
\subsection{Bounds on the period}
 We give some bounds on $N,$ that extend to general finite fields the results for odd characteristic prime fields of \cite[Th. 3]{R2}.
{\theorem \label{fonda} Keep the above notation.
\begin{enumerate}
\item If $P(a,b;x)$ is irreducible over $\F_q,$ then $N$ divides $(q+1)ord (-b).$
\item If $P(a,b;x)$ is reducible with two distinct roots over $\F_q,$ then $N$ divides $q-1$.
\item If $P(a,b;x)=(x-\alpha)^2$ is a square over $\F_q,$ a field of characteristic $p,$ then $N=p\, ord(\alpha).$
\end{enumerate}
}
\begin{proof}
We employ the expressions for $g_n$ derived in the three previous subsections.
\begin{enumerate}
\item Write $P(a,b;x)=(x-\alpha)(x-\alpha^q)=x^2-ax-b$ to get, identifying coefficients of $x^0,$ the relation $\alpha^{q+1}=-b.$ This implies that the order of $\alpha$ divides $(q+1)ord (-b).$
\item Write $P(a,b;x)=(x-\alpha)(x-\beta).$ By application of Fermat little theorem to $\F_q,$ the order of both $\alpha$ and $\beta$ divides $q-1.$
\item The sequence $n\mapsto (\lambda+\mu n)$ is periodic of period $p.$ The sequence $n\mapsto \alpha^n$ is periodic of order $ord(\alpha),$ a quantity dividing $q-1,$ and therefore coprime with $p.$
The sequence $n\mapsto (\lambda+\mu n)\alpha^n$ is therefore periodic of period $LCM(p,ord(\alpha))=p\,ord(\alpha).$
\end{enumerate}
\end{proof}
\section{Weight distribution}
{\theorem \label{dd} The dual distance of $C(a,b,q)$ is at least two and at most three. It is three when $x^2-ax-b$ has two distinct roots and when $K=1.$ In that case $C(a,b,q)$ is projective and  MDS.}
\begin{proof}
The distance of $C(a,b,q)^\bot,$ an $[N,N-2]$ is at most three by the Singleton bound. When it is three, the code $C(a,b,q)^\bot$ is MDS, and so is its dual by \cite[Chap 11, Th. 2]{MS}. Since the minimum distance of $C(a,b,q)$ is $N-K$ and its dimension $2,$ this happens iff $K=1.$  Since MDS codes have weights which are consecutive integers this cannot happen when $P(a,b;x)$ is a square by Theorem \ref{square}.
The dual distance is at least two by checking the generator matrix of $C(a,b,q)$ has no zero column. The generic column of the generator matrix can be seen to be
\begin{enumerate}
\item $(tr(\alpha^n),tr(\alpha^{n+1}))^t$ in the case $P(a,b;x)$ is irreducible ($tr(z)=z+z^q$).
\item $(\alpha^n,\beta^n)^t$ in the case $P(a,b;x)$ is reducible with two distinct roots.
\item $(\alpha^n,n\alpha^n)^t$ in the case $P(a,b;x)$ is a square.
\end{enumerate}
This proves the results.
\end{proof}

We compute the weight distribution when $C(a,b,q)$ is a two-weight code.

{\theorem If the code $C(a,b,q)$ is a two-weight code with weights $\{N-K,N\},$ then the respective frequencies are \{$\frac{(p-1)N}{K}, \frac{(p-1)(K(p+1)-N)}{K}$\}.
If the code $C(a,b,q)$ is a two-weight code with weights $\{N-\frac{N}{p},N\},$ then the respective frequencies are \{$p(p-1), p-1$\}.
}
\begin{proof} By Theorem \ref{dd} the dual distance of $C(a,b,q)$ is at least $2.$
 The result follows then by application of the Pless power moments and resolution of the system in the frequencies $x,y$ given by
\begin{eqnarray*}
x+y&=&p^2-1\\
x(N-K)+yN&=&p(p-1)N.
\end{eqnarray*}
The second assertion is obtained by replacing $K$ by $\frac{N}{p}$ in the first assertion.
\end{proof}

\section{Numerical examples}
We have avoided the examples with $a=b=1$ (Fibonacci recurrence) which are treated in \cite{SZS}. In the following tables, $r$ stands as a primitive root of $\F_q,$ as chosen in Magma \cite{Ma}.
\subsection{Irreducible cyclic codes}

\begin{center}
\small {Table 1} \\
\end{center}
\begin{center}
\small
\begin{tabular}{|c|c|c|c|c|c|c|c|c|c|c|}
  \hline
   $q$                  & 9       & 49      & 49      & 49       & 49       & 49        & 49        & 27      & 27      & 27 \\
  \hline $a$          & $r^{2}$ & $r^{5}$ & $r^{2}$ & $r^{10}$ & $r^{5 }$ & $r^{7}$   & $r^{27}$  & $r^{3}$ & $r^{7}$  & $r^{2}$ \\
  \hline $b$          & $r^{3}$ & $r^{30}$& $r^{15}$& $r^{9}$  & $r^{21}$ & $r^{11}$  & $r^{23}$  & $r^{11}$& $r^{14}$& $r^{10}$ \\
  \hline $N$          & 80      & 400     & 800     & 800      & 800      & 2400      & 2400      & 364     & 104     & 728 \\
  \hline $e$          & 10      & 25      & 50      & 50       & 50       & 50        & 50        & 14      & 4       & 28    \\
  \hline
\end{tabular}
\end{center}

\subsection{Reducible cyclic codes}
\subsubsection{Two roots}
\begin{center}
\small {Table 2} \\
\end{center}
\begin{center}
\small
\begin{tabular}{|c|c|c|c|c|c|c|c|c|c|c|}
  \hline
   $q$                  & 9       & 49      & 49      & 49      & 49       & 49        & 49       & 121     & 121      & 121 \\
  \hline $a$          & $r^{4}$ & $r^{7}$ & $r^{4}$ & $r^{4}$ & $r^{4 }$ & $r^{4}$   & $r^{4}$  & $r^{14}$& $r^{22}$ & $r^{25}$ \\
  \hline $b$          & $r^{8}$ & $r^{8}$ & $r^{8}$ & $r^{11}$& $r^{13}$ & $r^{17}$  & $r^{23}$ & $r^{7}$ & $r^{7}$  & $r^{7}$ \\
  \hline $N$          & 8       & 24      & 48      & 48      & 48       & 48        & 48       & 120     & 120      & 120  \\
  \hline $e$          & 4       & 12      & 8       & 48      & 48       & 48        & 48       & 120     & 120      & 120  \\
  \hline
\end{tabular}
\end{center}
\pagebreak
\subsubsection{Single root}
\begin{center}
\small {Table 3} \\
\end{center}
\begin{center}
\small
\begin{tabular}{|c|c|c|c|c|c|c|c|c|c|c|}
  \hline
   $q $               & 9       & 49      & 49      & 49       & 49       & 49        & 49        & 121     & 121      & 121 \\
  \hline $N $         & 6      & 168     & 336     & 336      & 168      & 336       & 336        & 440     & 1320     & 132  \\
  \hline $a$          & $r^{8}$ & $r^{18}$& $r^{21}$& $r^{23}$ & $r^{26}$ & $r^{29}$  & $r^{33}$  & $r^{15}$& $r^{19}$ & $r^{22}$ \\
  \hline $b $         & $r^{4}$ & $r^{28}$& $r^{34}$& $r^{38}$ & $r^{44}$ & $r^{2} $  & $r^{10}$  & $r^{66}$& $r^{24}$ & $r^{80}$ \\

  \hline
\end{tabular}
\end{center}

\section{Conclusion and open problems}
In the present paper, we have proved the surprizing result that any cyclic code of dimension $2$  has at most two nonzero weights, by using in an essential way the properties of second order recurrences over finite fields.
It would be worthwhile but probably very difficult to extend this result to cyclic codes of higher dimensions.

\end{document}